`

# Quantum Brownian transport in a correlated Gaussian force

Yun Jeong Kang [a], Seungsik Min [b], Sung Kyu Seo [c], and Kyungsik Kim [d,*]

[a] *School of Liberal Studies, Wonkwang University, Iksan 54538, Republic of Korea*

[b] *Department of Natural Science, Korea Naval Academy, Changwon 51698, Republic of Korea*

[c] *Haena Ltd., Jejudo 63568, Republic of Korea*

[d] *Department of Physics, Pukyong National University, Busan 48513, Republic of Korea*

**Abstract**

We study a classical system-bath model in which a system particle is linearly coupled to a bath of harmonic oscillators. In a system subject to white and correlated Gaussian noises, we derive an expression for the joint probability density, and the mean squared values of the system-bath particles are calculated. For white noise, the mean squared values of the system-bath particles exhibit an anomalous time dependence, which is different from that of normal diffusion. In particular, for a correlated Gaussian noise, the mean squared displacement and mean squared velocity of the bath particle show superspreading growths of $t^5$ and $t^3$ in $t \ll \tau$, respectively. When $\tau = 0$, the mean squared velocity of a quantum particle under random noise is proportional to $t$, while the mean squared velocity of a bath particle in the presence of correlated Gaussian noise increases in proportion to $t^2$. This anomalous transport phenomenon results from the mixed derivative structure of the master equation, which couples with the transport coordinates in the diffusion dynamics of the relative coordinates. This result shows that the removal of dissipation in the Caldeira-Leggett framework leads to fundamentally different transport mechanisms characterized by non-diffusive quantum diffusion.

* Corresponding authors. E-mail: kskim@pknu.ac.kr

## 1.Introduction

The dynamics of quantum particles interacting with an external environment has been an important subject in nonequilibrium statistical physics and quantum transport theory [1-3]. Such systems provide a fundamental framework for understanding dissipation [4], decoherence, and transport processes in open quantum systems [5,6]. In many physical situations, the interaction between a particle and its thermal environment leads to diffusive motion and relaxation toward thermal equilibrium. One of the most widely used theoretical descriptions of open quantum systems [7,8] is the Caldeira–Leggett model [9,10], which describes the interaction of a quantum particle with a thermal reservoir through a master equation for the reduced density matrix. The Lindblad master equation [11] used in quantum optics is strictly valid at high temperature limits [3] and can be derived for the quantum Brownian motion from this master equation. On the other hand, in the case of zero-point oscillations at low temperatures, it does not exhibit quantum characteristics and may violate physical principles. Further research is needed to find a direct analytical and computational framework by utilizing the unique structure of the Caldeira-Leggett model in relation to the Lindblad master equation.

On the other hand, Kondaurov and Polyakov [12] analyzed changes in decoherence and transport diffusion using numerical simulation in a Caldeira–Leggett bath oscillator model with environmental damping. This is particularly a recent paper in which the two authors numerically studied the system and bath oscillator dynamics. Buxton et al. [13] have introduced the Caldirola-Kanai Lagrangian into the Caldeira-Leggett model to describe an environment of damped harmonic oscillators, and analyzed numerically the quantum Brownian motion that emerges from the decoherence and well-transfer processes. Santos and Menon [14] proposed two diffusion models for scaled Brownian motion with random diffusion coefficients governed by a probability equation. The first model was generalized and demonstrated through computer simulation that diffusion is non-Gaussian, and the second model was a generalization of the minimum diffusion coefficient model and was said to be suitable for explaining the transition phenomenon from a non-Gaussian process to a standard Gaussian process. In the diffusion motion of swimming bodies, Olsen et al. [15] discussed superballistic motion in a model of micro-

`

swimming bodies passing through a porous medium. They showed that the motion occurs as a direct result of the nutrient environment, which changes dynamically depending on the swimming body density. Flekkøy et al. [16] showed that superdiffusion occurs in the Markov description of particle interactions, and that the conservation concentration of particles for particle interactions is governed by Fick's law. They showed that the superballistic diffusion for the dimension $d \geq 2$ can occur when particle interactions are fully described at local concentrations. Recent studies have considered various extensions of the Caldeira–Leggett equation, including the effects of correlated noise, non-Markovian environments, and external potentials [17-19]. These studies have revealed that environmental correlations may lead to anomalous transport properties that differ significantly from standard diffusion. However, the role of dissipation in determining the transport mechanism remains an important question, particularly when the dissipative term is absent or suppressed.

In this work, we investigate the dynamics of a quantum particle governed by the Caldeira–Leggett equation in the absence of the dissipative term. The system is driven by a correlated Gaussian noise and subjected to an external harmonic potential. By introducing characteristic coordinates and applying Fourier–Laplace transforms, we obtain analytic solutions for the reduced density matrix and evaluate the mean square displacement of the particle. Our results show that the transport properties of the system differ fundamentally from those of the standard Caldeira–Leggett dynamics. The present results demon persistent super-ballistic quantum spreading driven by correlated state that removing dissipation in the Caldeira–Leggett framework leads to a qualitatively differ and transport mechanism. Instead of approaching normal diffusion, the system exhibits environmental noise. The organization of this paper is structured as follows: In Section 2, we introduce diffusive quantum Brownian motion in Caldeira–Leggett models with a random noise, and analyze the joint probability density and its mean squared displacement and velocity for the quantum and bath particles. In Section 3, we present the joint probability densities of Caldeira–Leggett model for the quantum and bath particles with a correlated Gaussian noise in the limits of $t \ll \tau$, $t \gg \tau$ and for $\tau = 0$. Section 4 discusses the extension to statistical quantities, and our conclusions are finally summarized in Section 5.

## 2. Quantum Brownian motion in classical Caldeira–Leggett model with a random noise

The quantum Brownian motion described by the Caldeira–Leggett model can reproduce classical stochastic dynamics in the high-temperature limit. In this section, we derive analytically the joint probability density for a Brownian particle starting from the Langevin equation with a random force.

The system–bath model consists of a system particle of mass $m$ linearly coupled to a bath of harmonic oscillators. The Hamiltonian of the total system is

$$H = \frac{p^2}{2m} + \sum_{i=1}^{N} [\frac{p_i^2}{2m_i} + \frac{1}{2} m_i \omega_i^2 \left( x_i - \frac{c_i}{m_i \omega_i^2} x \right)^2]. \tag{1}$$

This Hamiltonian represents the standard Caldeira–Leggett model, originally introduced by Caldeira and Leggett [9,10], where the system coordinate $x$ is bilinearly coupled to bath oscillators $x_i$ for $i = N$.

The first term constitutes the Hamiltonian (only kinetic energy) of an isolated particle. The second term is the Hamiltonian of the isolated bath. The third term describes the linear coupling system-bath interaction, and the fourth bilinear coupling term is necessary to ensure that there are no additional forces acting on the free particle at rest [17,19], and this one clearly vanishes when the Brownian particle does not couple to the oscillators in $k_B T \gg \hbar\omega$ . In the Caldeira–Leggett model, the noise correlation [12] is given by $\langle \xi(t)\xi(t') \rangle = \int_0^\infty d\omega J(\omega) \coth\left(\frac{\hbar\omega}{2k_B T}\right) \cos\omega(t - t')$, where $\coth(\hbar\omega/2k_B T) \approx \frac{2k_B T}{\hbar\omega}$ in the high temperature of the classical limit. Therefore, in this paper, since it becomes a random noise $\langle \xi(t)\xi(t') \rangle = 2m\gamma k_B T\delta(t)$.

For a free Brownian particle $(V(x) = 0)$, the classical equations of motion derived from Hamilton's equations are

$$m\ddot{x}(t) = -\int_0^t M(t - t')\dot{x}(t')dt' + \xi(t), \tag{2}$$

$$m_i \ddot{x}_i(t) = -m_i \omega_i^2 x_i - c_i x. \tag{3}$$

where $M(t) = \sum_{i=1}^{N} \frac{c_i^2}{m_i \omega_i^2} \cos(\omega_i t)$ is the memory kernel of the bath. The random force $\xi(t)$ represents thermal fluctuations with zero mean $\langle \xi(t) \rangle = 0$ and correlation $\langle \xi(t)\xi(t') \rangle = 2D\delta(t - t')$, which corresponds to a Gaussian white noise and $D = m\gamma k_B T$.

Now, we introduce the joint probability density $p(x, v, t)$ and $p(x_i, v_i, t)$ of finding a quantum particle and a bath particle in displacement and velocity $x$ , $v$ and $x_i$, $v_i$ as

`

$$p(x,v,t)=<\delta\big(x-x(t)\big)\delta\big(v-v(t)\big)>,\ p(x_i,v_i,t)=<\delta\big(x_i-x_i(t)\big)\delta\big(v_i-v_i(t)\big)> \tag{4}$$

subject to the initial condition $p(x,v,0)=<\delta(x-x_0)\delta(v-v_0)>$ and $p(x_i,v_i,0)=<\delta(x_i-x_{i0})\delta(v_i-v_{i0})>$. The time evolution equations for $p(x,v,t)$ and $p(x_i,v_i,t)$ are given by

$$\frac{\partial}{\partial t}p(x,v,t)=-\frac{\partial}{\partial x}<\frac{\partial x}{\partial t}\delta\big(x-x(t)\big)\delta\big(v-v(t)\big)>-\frac{\partial}{\partial v}<\frac{\partial v}{\partial t}\delta\big(x-x(t)\big)\delta\big(v-v(t)\big)>, \tag{5}$$

$$\frac{\partial}{\partial t}p\big(x_i,v_{x_i},t\big)=-\frac{\partial}{\partial x_i}<\frac{\partial x_i}{\partial t}\delta\big(x_i-x_i(t)\big)\delta\big(v_i-v_i(t)\big)>-\frac{\partial}{\partial v_i}<\frac{\partial v_i}{\partial t}\,\delta(x_i-x_i(t))\delta(v_i-v_i(t))>. \tag{6}$$

Inserting Eq. (2) and Eq. (3) into Eq. (5) and Eq. (6) and manipulating the method of Ref. [20], two Fokker-Plank equations for $p(x,v,t)$ and $p(x_i,v_i,t)$ are derived as

$$\frac{\partial}{\partial t}p(x,v,t)=[-v\frac{\partial}{\partial x}+\frac{c_i^2}{\omega_i^2}v\frac{\partial}{\partial v}+\frac{x_0c_i^2}{\omega_i^2}cos\omega_i t\frac{\partial}{\partial v}]p(x,v,t)+D\frac{\partial^2}{\partial v^2}p(x,v,t), \tag{7}$$

$$\frac{\partial}{\partial t}p(x_i,v_i,t)=[-v_i\frac{\partial}{\partial x_i}+[k_ix_i-c_i(x_0+v_0)t+(t-1)\frac{c_i^3}{\omega_i^2}]\frac{\partial}{\partial v_i}]p(x_i,v_i,t)++c_iDt\frac{\partial^2}{\partial v_i^2}]p(x_i,v_i,t). \tag{8}$$

where the parameter $k_i$ are given by $k_1=\omega_i^2$ and the dimensionless $m=m_1=1$, respectively. The double Fourier transform of the joint probability density is defined by

$$p(\zeta,\nu,t)=\int_{-\infty}^{+\infty}dx\int_{-\infty}^{+\infty}dv\,exp(-i\zeta x-i\nu v)\,p(x,v,t), \tag{9}$$

$$p(\zeta_i,\nu_i,t)=\int_{-\infty}^{+\infty}dx_i\int_{-\infty}^{+\infty}dv_i\,exp(-i\zeta_ix_i-i\nu_iv_i)\,p(x_i,v_i,t). \tag{10}$$

By using the double Fourier transform in Eq. (7) and Eq. (8), we derive the Fourier transform of two Fokker-Planck equations for a quantum particle as

$$\frac{\partial}{\partial t}p(\zeta,\nu,t)=\left[\zeta-\frac{c_i^3}{\omega_i^2}\nu\right]\frac{\partial}{\partial\nu}p(\zeta,\nu,t)-icos\omega_i t\frac{x_0c_i^2}{\omega_i^2}\nu p(\zeta,\nu,t)-D\nu^2p(\zeta,\nu,t)+Ap(\zeta,\nu,t), \tag{11}$$

$$\frac{\partial}{\partial t}p(\zeta_i,\nu_i,t)=\left[-k_i\nu_i\frac{\partial}{\partial\zeta_i}+\zeta_i\frac{\partial}{\partial\nu_i}\right]p(\zeta_i,\nu_i,t)+i\left[c_i(x_0+v_0)t+(t-1)\frac{c_i^3}{\omega_i^2}\right]\nu_i-[c_iDt\nu_i^2p(\zeta,\nu,t)+A]p(\zeta_i,\nu_i,t). \tag{12}$$

Here, *A* is the separation constant. The second term on the right side of Eq. (11) is an imaginary part, so it is excluded from the calculation of the real part. Consequently, we get two Fokker-Planck equations, Eq. (7) and Eq. (8), and the Fourier transforms of Fokker-Planck equations, Eq. (11) and Eq. (12).

From now on, we obtain the joint probability density for a quantum particle with a random noise in the short- and long-time domains. When the steady state $\frac{\partial}{\partial t}p(\zeta,\nu,t)=0$, we have $p(\zeta,\nu,t)\to p^{st}(\zeta,\nu,t)$. $p^{st}(\zeta,\nu,t)$ is obtained as

$$p^{st}(\zeta,\nu,t)=exp\left[\frac{D}{(\zeta-c_i^3\nu/\omega_i^2)}\frac{\nu^3}{3}\right]. \tag{13}$$

At next step, the probability density $p(\zeta,\nu,t)$ is given by

$$p(\zeta,\nu,t)=q(\zeta,\nu,t)p^{st}(\zeta,\nu,t). \tag{14}$$

The time derivative of the probability density $q(\zeta,\nu,t)$ is given by $\frac{\partial}{\partial t}q(\zeta,\nu,t)=(\zeta-c_i^3\nu/\omega_i^2)\frac{\partial}{\partial\nu}q(\zeta,\nu,t)$. An arbitrary function of the variable $t-\nu/(\zeta-c_i^3\nu/\omega_i^2)$ becomes $q(\zeta,\nu,t)=\Phi[t-\nu/(\zeta-c_i^3\nu/\omega_i^2)]$. Thus,

$$p(\zeta,\nu,t)=q(\zeta,\nu,t)p^{st}(\zeta,\nu,t)=\Phi[t-\nu/(\zeta-c_i^3\nu/\omega_i^2)]exp\left[\frac{D}{(\zeta-c_i^3\nu/\omega_i^2)}\frac{\nu^3}{3}\right]. \tag{15}$$

From Eq. (15), the Fourier transform of the probability density is calculated as

$$p(\zeta,\nu,t)=exp\left[-\frac{D}{3}t^3\zeta^2-\frac{Dc_i^6}{3\omega_i^4}t^3\nu^2\right]. \tag{16}$$

Using the inverse Fourier transform, the probability densities become

`

$$p(x,t) = [4\pi Dt^3/3]^{-\frac{1}{2}} exp[-\frac{3x^2}{4Dt^3}], p(v,t) = [4\pi Dc_i^6 t^3/3\omega_i^4]^{-\frac{1}{2}} exp[-\frac{3\omega_i^4}{4Dc_i^6 t^3} v^2]. \quad (17)$$

Thus, the mean squared quantities corresponding to Eq. (17) are given by

$$< x^2(t) >= \frac{2D}{3} t^3, \ < v^2(t) >= \frac{2Dc_i^6}{3\omega_i^4} t^3. \quad (18)$$

in all time (short- and long-time) domains.

Next, we obtain the joint probability density for a $i-th$ bath particle with a random noise in the short- and long-time domains. Separating the variables as $p(\zeta_i, v_i, t) = p(\zeta_i, t) p(v_i, t)$ from Eq. (12), we have

$$\frac{\partial}{\partial t} p(\zeta_i, t) = -k_i v_i \frac{\partial}{\partial \zeta_i} - \frac{1}{2}[c_i D t v_i^2 + A + 2B] p(\zeta_i, t), \quad (19)$$

$$\frac{\partial}{\partial t} p(v_i, t) \ = \zeta_i \frac{\partial}{\partial v_i} p(v_i, t) - \frac{1}{2}[c_i D t v_i^2 + A - 2B] p(v_i, t). \quad (20)$$

Here, $A$ and $B$ denote the separation constants. When two probability densities are in a steady state, we can consider the situation as follows: As $\frac{\partial}{\partial t} p(\zeta_i, t) = 0$, $\frac{\partial}{\partial t} p(v_i, t) = 0$, we have $p(\zeta_i, t) \to p^{st}(\zeta_i, t)$ and $p(v_i, t) \to p^{st}(v_i, t)$ for $\zeta_i$ and $v_i$. Then, Eqs. (19) and (20) are given by

$$-k_i v_i \frac{\partial}{\partial \zeta_i} p^{st}(\zeta_i, t) - \frac{1}{2}[c_i D t v_i^2 + A + 2B] p^{st}(\zeta_i, t) = 0, \quad (21)$$

$$\zeta_i \frac{\partial}{\partial v_i} p^{st}(v_i, t) \ - \frac{1}{2}[c_i D t v_i^2 + A - 2B] p^{st}(v_i, t) \ = 0. \quad (22)$$

The steady Fourier transform of the probability densities $p^{st}(\zeta_i, t)$ and $p^{st}(v_i, t)$ is calculated by

$$p^{st}(\zeta_i, t) = exp[-\frac{Dc_i t}{2k_i v_i} \zeta_i v_i^2 - \frac{A+2B}{k_i v_i} \zeta_i], \quad (23)$$

$$p^{st}(v_i, t) = exp[\frac{Dc_i t}{6\zeta_i} v_i^3 - \frac{A-2B}{2\zeta_i} v_i]. \quad (24)$$

At next step, the probability densities $p(\zeta_i, t)$ and $p(v_i, t)$ are given by

$$p(\zeta_i, t) = q(\zeta_i, t) p^{st}(\zeta_i, t), \ p(v_i, t) = q(v_i, t) p^{st}(v_i, t). \quad (25)$$

The time derivatives of the probability density $q(\zeta_i, t)$ and $q(v_i, t)$ are given by $\frac{\partial}{\partial t} q(\zeta_i, t) = -k_i v_i \frac{\partial}{\partial \zeta_i} q(\zeta_i, t)$ and $\frac{\partial}{\partial t} q(v_i, t) = \zeta_i \frac{\partial}{\partial v_i} q(v_i, t)$, respectively. The arbitrary functions of the variable $t - \zeta_i / k_i v_i$ and $t + v_i / \zeta_i$ become $q(\zeta_i, t) = \Phi[t - \zeta_i / k_i v_i]$ and $q(v_{i_i}, t) = \Phi[t + v_i / \zeta_i]$. Thus, we obtain

$$p(\zeta_i, t) = q(\zeta_i, t) p^{st}(\zeta_i, t) = \Phi[t - \zeta_i / k_i v_i] \ exp[-\frac{Dc_i t}{2k_i v_i} \zeta_i v_i^2 - \frac{A+2B}{k_i v_i} \zeta_i], \quad (26)$$

$$p(v_i, t) = q(v_i, t) p^{st}(v_i, t) = \Phi[t + v_i / \zeta_i] \ exp[\frac{Dc_i t}{6\zeta_i} v_i^3 - \frac{A-2B}{2\zeta_i} v_i]. \quad (27)$$

Then we obtain $p(\zeta_i, v_i, t) = p(\zeta_i, t) p(v_i, t)$ as

$$p(\zeta_i, v_i, t) = p(\zeta_i, t) p(v_i, t) = exp\left[-\frac{Dc_i}{6} t^4 \zeta_i^2 - \frac{Dc_i}{2} t^2 v_i^2\right]. \quad (28)$$

Using the inverse Fourier transform, the probability densities are given by

$$p(x_i, t) = [2\pi Dc_i t^4/3]^{-\frac{1}{2}} exp[-\frac{3}{2Dc_i t^4} x_i^2], p(v_i, t) = [2\pi Dc_i t^2]^{-\frac{1}{2}} exp[-\frac{1}{2Dc_i t^2} v_i^2]. \quad (29)$$

with the mean square displacement and velocity

$$< x_i^2(t) >= \frac{Dc_i}{3} t^4, \ < v_i^2(t) >= Dc_i t^2. \quad (30)$$

## 3. Quantum Brownian motion in classical Caldeira–Leggett model with a correlated Gaussian noise

The classical equations of motion for a quantum particle and a $i-th$ bath particle with a corelated Gaussian noise $\eta(t)$ are

`

$$m\ddot{x}(t) = -\int_0^t M(t-t')\dot{x}(t')dt' + \eta(t), \quad (31)$$

$$m_i\ddot{x}_i(t) = -m_i\omega_i^2 x_i - c_i x. \quad (32)$$

where $M(t) = \sum_i \frac{c_i^2}{m_i\omega_i^2}\cos(\omega_i t)$. The corelated Gaussian force $\eta(t)$ represents thermal fluctuations with zero mean $\langle\eta(t)\rangle = 0$ and correlation

$$\langle\eta(t)\eta(t')\rangle = \frac{D}{\tau}exp(-\frac{|t-t'|}{\tau}). \quad (33)$$

which corresponds to a correlated Gaussian noise and $\tau$ is the correlation time.

We introduce the time evolution equations of the joint probability densities of the particle position and velocity as Eqs. (5) and (6). Inserting Eq. (31) and Eq. (32) into Eq. (5) and Eq. (6) and manipulating the method of Ref. [20], two Fokker-Plank equations for $p(x,v,t)$ and $p(x_i,v_i,t)$ are derived as

$$\frac{\partial}{\partial t}p(x,v,t) = [-v\frac{\partial}{\partial x} + \frac{c_i^2}{\omega_i^2}v\frac{\partial}{\partial v} + \frac{x_0 c_i^2}{\omega_i^2}cos\omega_i t\frac{\partial}{\partial v}]p(x,v,t) + D[-b(t)\frac{\partial^2}{\partial x\partial v} + a(t)\frac{\partial^2}{\partial v^2}]p(x,v,t), \quad (34)$$

$$\frac{\partial}{\partial t}p(x_i,v_i,t) = [-v_i\frac{\partial}{\partial x_i} + [k_i x_i - c_i(x_0+v_0)t + (t-1)\frac{c_i^3}{\omega_i^2}]\frac{\partial}{\partial v_i}]p(x_i,v_i,t) + +c_i Dt[-b(t)\frac{\partial^2}{\partial x\partial v} + a(t)\frac{\partial^2}{\partial v^2}]p(x_i,v_i,t). \quad (35)$$

where the parameters $a(t) = 1-\exp(-t/\tau)$ and $b(t) = (t+\tau)\exp(-t/\tau) - \tau$. The third term on the right side of Eq. (34) is an imaginary part, so it is excluded from the calculation of the real part.

By using the double Fourier transform in Eq. (34) and Eq. (35), we derive the Fourier transform of two Fokker-Planck equations for a quantum particle as

$$\frac{\partial}{\partial t}p(\zeta,v,t) = \left[\zeta - \frac{c_i^3}{\omega_i^2}v\right]\frac{\partial}{\partial v}p(\zeta,v,t) + D[b(t)\zeta v - a(t)v^2]p(\zeta,v,t) + Ep(\zeta,v,t), \quad (36)$$

$$\frac{\partial}{\partial t}p(\zeta_i,v_i,t) = \left[-k_i v_i\frac{\partial}{\partial\zeta_i} + \zeta_i\frac{\partial}{\partial v_i}\right]p(\zeta_i,v_i,t) + i\left[c_i(x_0+v_0)t + (t-1)\frac{c_i^3}{\omega_i^2}\right]v_i p(\zeta_i,v_i,t)$$
$$+c_i Dt\left[b(t)\zeta_i v_i - a(t)v_i^2\right]p(\zeta,v,t) - Ep(\zeta_i,v_i,t). \quad (37)$$

### *3.1. Probability densities* $p(x,t)$ *and* $p(v,t)$ *for a quantum particle in three-time domains*

In this subsection, we obtain the probability densities $p(x,t)$ and $p(v,t)$ for a quantum particle in the limits of $t \ll \tau$, $t \gg \tau$ and for $\tau = 0$.

3.1.1. $p(x,t)$ and $p(v,t)$ in time domain $t \ll \tau$

By neglecting an imaginary term in the right-hand side of Eq. (36), Eq. (36) becomes

$$\frac{\partial}{\partial t}p(\zeta,v,t) = \left[\zeta - \frac{c_i^3}{\omega_i^2}v\right]\frac{\partial}{\partial v}p(\zeta,v,t) + D[b(t)\zeta v - a(t)v^2]p(\zeta,v,t) + Ep(\zeta,v,t). \quad (38)$$

$p(\zeta,v,t)$ becomes $p^{st}(\zeta,v,t)$ by taking $\frac{\partial}{\partial t}p(\zeta,v,t) = 0$ for $\zeta_1$ in steady state. The Fourier-transformed equations for the steady joint probability density separate as

$$\left[\zeta - \frac{c_i^3}{\omega_i^2}v\right]\frac{\partial}{\partial v}p^{st}(\zeta,v,t) + D[b(t)\zeta v - a(t)v^2]p^{st}(\zeta,v,t) + Ep^{st}(\zeta,v,t) = 0. \quad (39)$$

The stationary solution from Eq. (39) is

$$p^{st}(\zeta,v,t) = exp[\frac{1}{(\zeta - c_i^3 v/\omega_i^2)}[D(-b(t)\zeta\frac{v^2}{2} + a(t)\frac{v^3}{3}) + Ev]]. \quad (40)$$

A factorization is introduced: $p(\zeta_1,t) \equiv q(\zeta,v,t)p^{st}(\zeta,v,t)$. Then, $p(\zeta,v,t)$ is calculated as

$$p(\zeta,v,t) = q(\zeta,v,t)\,exp[\frac{1}{\left(\zeta - \frac{c_i^3 v}{\omega_i^2}\right)}[D(-b(t)\zeta\frac{v^2}{2} + a(t)\frac{v^3}{3}) + Ev]]. \quad (41)$$

Repeating this transformation successively, we obtain

`

$$q(\zeta,v,t) = r(\zeta,v,t)\, exp[\frac{D}{\left(\zeta-\frac{c_i^3 v}{\omega_i^2}\right)^2}[-b^{'(t)}\zeta\frac{v^3}{6} + a'(t)\frac{v^4}{12}]], \tag{42}$$

$$r(\zeta,v,t) = s(\zeta,v,t)\, exp[\frac{D}{\left(\zeta-\frac{c_i^3 v}{\omega_i^2}\right)^3}[-b^{''(t)}\zeta\frac{v^4}{24} + a''(t)\frac{v^5}{60}]], \tag{43}$$

$$s(\zeta,v,t) = t(\zeta,v,t)\, exp[-\frac{D}{\left(\zeta-\frac{c_i^3 v}{\omega_i^2}\right)^4}[b'''(t)\zeta\frac{v^5}{120}]]. \tag{44}$$

We have $p(\zeta,v,t)$ from Eqs. (41)-(44) as

$$p(\zeta,v,t) = t(\zeta,v,t)s^{st}(\zeta,v,t)r^{st}(\zeta,v,t)q^{st}(\zeta,v,t)p^{st}(\zeta,v,t)$$

$$= \Theta\left[t + \frac{v}{(\zeta - c_i^3 v/\omega_i^2)}\right] s^{st}(\zeta,v,t)r^{st}(\zeta,v,t)q^{st}(\zeta,v,t)p^{st}(\zeta,v,t). \tag{45}$$

Here, $a'(t) = da(t)/dt$ and $a''(t) = d^2a(t)/dt^2$. Neglecting the terms proportional to $1/\tau^3$, $t(\zeta,v,t)$ is given by $\frac{\partial}{\partial t}t(\zeta,v,t) = (\zeta - c_i^3 v/\omega_i^2)\frac{\partial}{\partial \zeta_1}t(\zeta,v,t)$. An arbitrary function of the combination $t + v/(\zeta - c_i^3 v/\omega_i^2)$ becomes $t(\zeta,v,t) = \Theta[t + v/(\zeta - c_i^3 v/\omega_i^2)]$. Expanding derivatives to the second order in powers of $t/\tau$ and performing cancelations yields

$$p(\zeta,v,t) = exp[-\frac{Dt^4}{8\tau}\zeta^2 - \frac{Dc_i^3 t^3}{\tau\omega_i^2}v^2]. \tag{46}$$

Using the inverse Fourier transform, the spatial probability densities $p(x,t)$ and $p(v,t)$ are

$$p(x,t) = [\pi\frac{Dt^4}{2\tau}]^{-1/2}exp[-\frac{2\tau}{Dt^4}x^2,\ p(v,t) = [4\pi\frac{Dc_i^3 t^3}{\tau\omega_i^2}]^{-\frac{1}{2}}exp[-\frac{\tau\omega_i^2}{4Dc_i^3 t^3}v^2]. \tag{47}$$

with the mean square values

$$< x^2(t) >= \frac{D}{4\tau}t^4,\ < v^2(t) >= \frac{2Dc_i^3}{\tau\omega_i^2}t^3. \tag{48}$$

3.1.2. $p(x,t)$ and $p(v,t)$ in time domain $t \gg \tau$

In the long-time domain, we approximate Eq. (36) as

$$p_v(\zeta,v,t) \cong D[b(t)\zeta v - a(t)v^2]p_v(\zeta,v,t), \tag{49}$$

where $p(\zeta,v,t) \equiv p_v(\zeta,v,t)$. Thus,

$$p_v(\zeta,v,t) = exp[D\int[b(t)\zeta v - a(t)v^2]dt]. \tag{50}$$

Defining $q_v^{st}(\zeta,v,t)$ via $q_v(\zeta,v,t)p_v(\zeta,v,t)$, we obtain

$$q_v^{st}(\zeta,v,t) = exp[-D\int[b(t)\zeta v - a(t)v^2]dt]. \tag{51}$$

From Eq. (40), the stationary probability density $p^{st}(\zeta,v,t)$ is

$$p^{st}(\zeta,v,t) = exp[\frac{1}{(\zeta - c_i^3 v/\omega_i^2)}[D(b(t)\zeta\frac{v^2}{2} - a(t)\frac{v^3}{3}) + Ev]]. \tag{52}$$

Taking arbitrary functions of $t - v/(\zeta - c_i^3 v/\omega_i^2)$, i.e., $r(\zeta_1,t) = \Theta[t + v/(\zeta - c_i^3 v/\omega_i^2)]$, and expanding in powers of $t/\tau$ yields the following after cancelations:

$$p(\zeta,v,t) = r(\zeta,v,t)q_v^{st}(\zeta,v,t)p^{st}(\zeta,v,t) = \Theta[t + v/(\zeta - c_i^3 v/\omega_i^2)]q_v^{st}(\zeta,v,t)p^{st}(\zeta,v,t)$$

$$= exp[-\frac{Dt^3}{3}\zeta^2 - \frac{Dc_i^3 t^3}{\tau\omega_i^2}v^2], \tag{53}$$

where $\int a(t)dt = t - \tau$ (with $a(t) = 1$) and $\int b(t)dt = -\tau t$, (with $b(t) = -\tau$) in the long-time domain. Using the inverse Fourier transform, the long-time densities are

$$p(x,t) = [4\pi Dt^3/3]^{-1/2}exp[-\frac{3}{4Dt^3}x^2,\ p(v,t) = [4\pi\frac{Dc_i^3 t^3}{\tau\omega_i^2}]^{-\frac{1}{2}}exp[-\frac{\tau\omega_i^2}{4Dc_i^3 t^3}v^2]. \tag{54}$$

Thus, the mean squared quantities for $p(x,t)$ and $p(v,t)$ from Eq. (54) are, respectively, given by

`

$$< x^2(t) >= \frac{2D}{3} t^3, \; < v^2(t) >= \frac{2Dc_i^3}{\tau\omega_i^2} t^3. \tag{55}$$

3.1.3. $p(x,t)$ and $p(v,t)$ for $\tau = 0$

For $\tau = 0$, Eq. (36) reduces to

$$\frac{\partial}{\partial t} p(\zeta,v,t) = \left[\zeta - \frac{c_i^3}{\omega_i^2} v\right] \frac{\partial}{\partial v} p(\zeta,v,t) - Dv^2 p(\zeta,v,t), \tag{56}$$

where $a(t) = 1$, and $b(t) = 0$. In steady state, we calculate $p^{st}(\zeta,v,t)$ as

$$p^{st}(\zeta,v,t) = exp[\frac{D}{(\zeta - c_i^3 v/\omega_i^2)} \frac{v^3}{3}]. \tag{57}$$

Using the approximation $[\zeta - \frac{c_i^3}{\omega_i^2} v]^{-1} \cong \zeta^{-1}[1 - \frac{c_i^3}{\omega_i^2} v\zeta^{-1}]$ in Eq. (57), the dependent forms are

$$p(\zeta,v,t) = \Theta\left[t - v/\left(\zeta - \frac{c_i^3}{\omega_i^2} v\right)\right] p^{st}(\zeta,v,t). \tag{58}$$

Therefore,

$$p(\zeta,v,t) = \exp[-\frac{1}{3} Dt^3\zeta^2 - Dtv^2]. \tag{59}$$

Inverse Fourier transforms give

$$p(x,t) = [4\pi Dt^3/3]^{-1/2} exp[-\frac{3}{4Dt^3} x^2, \; p(v,t) = [4\pi Dt]^{-\frac{1}{2}} exp[-\frac{1}{4Dt} v^2]. \tag{60}$$

Thus, the mean squared deviations for $p(x,t)$ and $p(v,t)$ are

$$< x^2(t) >= \frac{2D}{3} t^3, \; < v^2(t) >= 2Dt. \tag{61}$$

### *3.2. Probability densities $p(x_i,t)$ and $p(v_i,t)$ for a bath particle in three-time domains*

In this subsection, we obtain the probability densities $p(x_i,t)$ and $p(v_i,t)$ for a bath particle in the limits of $t \ll \tau$, $t \gg \tau$ and for $\tau = 0$.

3.2.1. $p(x,t)$ and $p(v,t)$ in short-time domain $t \ll \tau$

From Eq. (37), by separating variables $\zeta_i$ and $v_i$ as $p(\zeta_i, v_i, t) = p(\zeta_i, t)p(v_i, t)$, the Fourier transforms of the probability densities $p(\zeta_i, t)$ and $p(v_i, t)$ become

$$\frac{\partial}{\partial t} p(\zeta_i,t) = -k_i v_i \frac{\partial}{\partial \zeta_i} p(\zeta_i,t) + \frac{1}{2} c_i Dt\left[b(t)\zeta_i v_i - a(t)v_i^2\right] p(\zeta_i,t) + (F - \frac{E}{2}) p(\zeta_i,t), \tag{62}$$

$$\frac{\partial}{\partial t} p(v_i,t) = \zeta_i \frac{\partial}{\partial v_i} p(v_i,t) + \frac{1}{2} c_i Dt\left[b(t)\zeta_i v_i - a(t)v_i^2\right] p(v_i,t) - (F + \frac{E}{2}) p(v_i,t). \tag{63}$$

Here, $F$ is the separation constant. By taking $\frac{\partial}{\partial t} p(\zeta_i,t) = 0$ for $\zeta_i$ in steady state, $p(\zeta_i,t)$ becomes $p^{st}(\zeta_i,t)$. The Fourier-transformed equations for the steady joint probability density separate as

$$-k_i v_i \frac{\partial}{\partial \zeta_i} p^{st}(\zeta_i,t) + \frac{1}{2} c_i Dt\left[b(t)\zeta_i v_i - a(t)v_i^2\right] p^{st}(\zeta_i,t) + \left(F - \frac{E}{2}\right) p^{st}(\zeta_i,t) = 0, \tag{64}$$

$$\zeta_i \frac{\partial}{\partial v_i} p^{st}(v_i,t) + \frac{1}{2} c_i Dt\left[b(t)\zeta_i v_i - a(t)v_i^2\right] p^{st}(v_i,t) - \left(F + \frac{E}{2}\right) p^{st}(v_i,t) = 0. \tag{65}$$

From (64), $p^{st}(\zeta_i,t)$ for $\zeta_i$ in the steady state is given by

$$p^{st}(\zeta_i,t) = exp[\frac{Dc_i t}{2k_i v_i}\left[b(t)\frac{\zeta_i^2}{2}\zeta_i v_i - a(t)\zeta_i v_i^2\right] - \frac{1}{k_i v_i}\left(F - \frac{E}{2}\right)\zeta_i]. \tag{66}$$

Similarly, it is expedient to perform the sequence of successive transformations as

$$p(\zeta_i,t) = q(\zeta_i,t)\, exp[\frac{Dc_i t}{2k_i v_i}\left[b(t)\frac{\zeta_i^2}{2} v_i - a(t)\zeta_i v_i^2\right] - \frac{1}{k_i v_i}\zeta, v\left(F - \frac{E}{2}\right)\zeta_i], \tag{67}$$

`

$$q(\zeta_i,t) = r(\zeta_i,t)\, exp[-\frac{Dc_it}{2(k_iv_i)^2}\Big[b'(t)\frac{\zeta_i^3}{6}v_i - a'(t)\frac{\zeta_i^2}{2}v_i^2\Big], \tag{68}$$

$$r(\zeta_i,t) = s(\zeta_i,t)\, exp[+\frac{Dc_it}{2(k_iv_i)^3}\Big[b''(t)\frac{\zeta_i^4}{24}v_i - a''(t)\frac{\zeta_i^3}{6}v_i^2\Big], \tag{69}$$

$$s(\zeta_i,t) = u(\zeta_i,t)\, exp[-\frac{Dc_it}{2(k_iv_i)^4}\Big[b'''(t)\frac{\zeta_i^5}{120}v_i\Big]]. \tag{70}$$

We have $p(\zeta_i,t)$ from Eqs. (66)-(70) as

$$p(\zeta_i,t) = u(\zeta_i,t)s^{st}(\zeta_i,t)r^{st}(\zeta_i,t)q^{st}(\zeta_i,t)p^{st}(\zeta_i,t) = \Theta\left[t - \frac{\zeta_i}{k_iv_i}\right]s^{st}(\zeta_i,t)r^{st}(\zeta_i,t)q^{st}(\zeta_i,t)p^{st}(\zeta_i,t). \tag{71}$$

Here, $a'(t) = da(t)/dt$ and $a''(t) = d^2a(t)/dt^2$. Neglecting the terms proportional to $1/\tau^3$, $u(\zeta_i,t)$ is given by $\frac{\partial}{\partial t}u(\zeta_i,t) = -k_iv_i\zeta_i\frac{\partial}{\partial\zeta_i}u(\zeta_i,t)$. An arbitrary function of the combination $t - \zeta_i/k_iv_i$ becomes $u(\zeta_i,t) = \Theta[t - \zeta_i/k_iv_i]$. Then, we have

$$p(\zeta_i,t) = exp[-\frac{Dt^4}{8\tau}\zeta_i^2 - \frac{Dc_i^3t^3}{\tau\omega_i^2}v_i^2]. \tag{72}$$

Using the inverse Fourier transform, the spatial probability densities $p(x,t)$ and $p(v,t)$ are

$$p(x_i,t) = [\pi\frac{Dt^4}{2\tau}]^{-\frac{1}{2}}exp[-\frac{2\tau}{Dt^4}x_i^2],\ p(v_i,t) = [4\pi\frac{Dc_i^3t^3}{\tau\omega_i^2}]^{-\frac{1}{2}}exp[-\frac{\tau\omega_i^2}{4Dc_i^3t^3}v_i^2]. \tag{73}$$

with the mean square values

$$< x_i^2(t) >= \frac{Dt^4}{4\tau}t^4,\ < v_i^2(t) >= \frac{2Dc_i^3}{\tau\omega_i^2}t^3. \tag{74}$$

Next, in the time domain $t \ll \tau$, From (64), $p^{st}(v_i,t)$ for $v_i$ in the steady state, from (65), is given by

$$p^{st}(v_i,t) = exp[\frac{Dc_it}{2\zeta_i}\Big[a(t)\frac{v_i^3}{3} - b(t)\zeta_i\frac{v_i^2}{2}\Big] - \frac{1}{\zeta_i}\Big(F + \frac{E}{2}\Big)v_i]. \tag{75}$$

It is now expedient to perform the sequence of successive transformations as

$$p(v_i,t) = q(v_i,t)p^{st}(v_i,t), \tag{76}$$

$$q(v_i,t) = r(v_i,t)\, exp[\frac{Dc_it}{2\zeta_i^2}[\frac{a'(t)}{12}v_i^4 - \frac{b'(t)}{6}\zeta_iv_i^3], \tag{77}$$

$$r(v_i,t) = s(v_i,t)\, exp[\frac{Dc_it}{2\zeta_i^3}[\frac{a''(t)}{60}v_i^5 - \frac{b''(t)}{24}\zeta_iv_i^4], \tag{78}$$

$$s(v_i,t) = u(v_i,t)\, exp[-\frac{Dc_it}{\zeta_i^4}\frac{b'''(t)}{240}\zeta_iv_i^5]. \tag{79}$$

Getting rid of the terms proportional to $1/\tau^3$, the Fourier transform of the probability density $Z(v_2,t)$ obeys $\frac{\partial}{\partial t}u(v_i,t) = \zeta_i\frac{\partial}{\partial v_i}u(v_i,t)$. Then, we satisfy the solution as a function of variable $t + v_i/\zeta_i$, and an arbitrary function is given by $u(v_i,t) = \Phi[t + v_i/\zeta_i]$. Expanding their derivatives to the second order in powers of $t/\tau$, we derive and calculate the expression for $p(v_i,t)$ after some cancellations as

$$p(v_i,t) = \Phi[t + v_i/\zeta_i]s^{st}(v_i,t)r^{st}(v_i,t)q^{st}(v_i,t)p^{st}(v_i,t). \tag{80}$$

We calculate from Eqs. (25) and (32) that

$$p(v_i,t) = exp[-\frac{Dc_it^5}{8\tau}\zeta_i^2 - \frac{Dc_it^3}{2\tau}v_i^2]. \tag{81}$$

The inverse Fourier transforms of $p(\zeta_i,t)$ and $p(v_i,t)$ for displacement $x_i$ and velocity $v_i$ yield

$$p(x_i,t) = [\pi\frac{Dc_it^5}{2\tau}]^{-1/2}\, exp[-\frac{2\tau}{Dc_it^5}x_i^2],\ p(v_i,t) = [2\pi\frac{Dc_it^3}{\tau}]^{-1/2}\, exp[-\frac{\tau}{2Dc_it^3}v_i^2]. \tag{82}$$

We get the mean squared displacement and the mean squared velocity as

$$< x_i^2(t) >= \frac{Dc_i}{4\tau}t^5,\ < v_i^2(t) >= \frac{Dc_i}{\tau}t^3. \tag{83}$$

Consequently, from Eqs. (74) and (83), the mean squared displacement and the mean squared velocity in the limit of $t \ll \tau$ is obtained as

`

$$< x_i^2(t) >= \frac{Dc_i}{4\tau} t^5,\ < v_i^2(t) >= \frac{Dc_i}{\tau}(1 + \frac{2c_i^2}{\omega_i^2})t^3. \quad (84)$$

Therefore, the $< x_i^2(t) >$ and $< v_i^2(t) >$ calculated from the probability densities $p(x_i, t)$ and $p(v_i, t)$ for a bath particle in the limits of $t \ll \tau$ scale as $\sim t^5$ and $\sim t^3$, respectively.

3.2.2. $p(x_i, t)$ and $p(v_i, t)$ in long-time domain $t \gg \tau$

In this subsection, we obtain the probability densities $p(x_i, t)$ and $p(v_i, t)$ in the long-time domain as

$$\frac{\partial}{\partial t} p(\zeta_i, t) \cong \frac{1}{2} Dc_i t\big[b(t)\zeta_i v_i - a(t)v_i^2\big]p(\zeta_i, t),\ \frac{\partial}{\partial t} p(v_i, t) \cong \frac{1}{2} Dc_i t\big[b(t)\zeta_i v_i - a(t)v_i^2\big]p(v_i, t). \quad (85)$$

From Eqs. (84) and (85), we briefly have

$$p_{\zeta_i}(\zeta_i, t) = exp[\frac{Dc_i t}{2} \int \big[b(t)\zeta_i v_i - a(t)v_i^2\big]dt].. \quad (86)$$

We find $q_{\zeta_i}^{st}(\zeta_i, t)$ for $\zeta_i$ from $p(\zeta_i, t) \equiv q_{\zeta_2}(\zeta_i, t)p_{\zeta_i}(\zeta_i, t)$ as

$$q_{\zeta_i}^{st}(\zeta_i, t) = exp[-\frac{Dc_i t}{2} \int \big[b(t)\zeta_i v_i - a(t)v_i^2\big]dt]. \quad (87)$$

Then, we calculate $p^{st}(\zeta_i, t)$ as

$$p^{st}(\zeta_i, t) = exp[\frac{Dc_i t}{2k_i v_i}\Big[b(t)\frac{\zeta_i^2}{2}\zeta_i v_i - a(t)\zeta_i v_i^2\Big] - \frac{1}{k_i v_i}\Big(F - \frac{E}{2}\Big)\zeta_i], \quad (88)$$

$$q_{\zeta_i}(\zeta_i, t) = r(\zeta_i, t)q_{\zeta_2}^{st}(\zeta_i, t) = r(\zeta_i, t)\, exp[-\frac{Dc_i t}{2} \int \big[b(t)\zeta_i v_i - a(t)v_i^2\big]dt], \quad (89)$$

where $\int a_2(t)dt = t - \tau$, $a_2(t) = 1$ and $\int b_2(t)dt = -\tau t$, $b_2(t) = -\tau$ in the long-time domain. Taking the solution as an arbitrary function of variable $t - \zeta_i/k_i v_i$, the arbitrary function $r(\zeta_i, t)$ becomes $r(\zeta_i, t) = \Phi[t - \frac{\zeta_i}{k_i v_i}]$. By expanding in powers of $t/\tau$, we derive and calculate the expression for $p^{st}(\zeta_i, t)$ after some cancellations, as

$$p(\zeta_i, t) = r(\zeta_i, t)q_{\zeta_i}^{st}(\zeta_i, t)p^{st}(\zeta_i, t) = \Phi[t - \frac{\zeta_i}{k_i v_i}]q_{\zeta_i}^{st}(\zeta_i, t)p^{st}(\zeta_i, t). \quad (90)$$

In the long-time domain for $v_i$ from Eq. (85), we write $p(v_i, t)$ by a similar procedure as Eqs. (86)-(90) for $\zeta_i$ as

$$p(v_i, t) = r(v_i, t)q_{v_i}^{st}(v_i, t)p^{st}(v_i, t) = \Phi[t + v_i/\zeta_i]q_{v_i}^{st}(v_{i_i}, t)p^{st}(v_i, t). \quad (91)$$

Therefore, from Eq. (90) and Eq. (91), we calculate that

$$p(\zeta_i, v_i, t) = p(\zeta_i, t)p(v_i, t) = exp[-\frac{Dc_i t^4}{6}\zeta_i^2 - \frac{Dc_i t^2}{2}v_i^2]. \quad (92)$$

From Eq. (91), using the inverse Fourier transform, the probability densities $p(x_i, t)$ and $p(v_i, t)$ are, respectively, presented by

$$p(x_i, t) = [2\pi \frac{Dc_i t^4}{3}]^{-1/2} exp[-\frac{3}{2Dc_i t^4}x_i^2],\ p(v_i, t) = [2\pi Dc_i t^2]^{-\frac{1}{2}} exp[-\frac{1}{2Dc_i t^2}v_i^2]. \quad (93)$$

The mean squared displacement and the mean squared velocity for $p(x_i, t)$ and $p(v_i, t)$ are, respectively, given by

$$< x_i^2(t) >= \frac{1}{3} Dc_i t^4,\ < v_i^2(t) >= Dc_i t^2. \quad (94)$$

3.2.3. $p(x_i, t)$ and $p(v_i, t)$ in time domain $\tau = 0$

For $\tau = 0$, we can write the ap proximate equations from Eqs. (23) and (24) for $\zeta_2$ and $v_2$ as

$$\frac{\partial}{\partial t} p(\zeta_i, t) = -k_i v_i \frac{\partial}{\partial \zeta_i} p(\zeta_i, t) - \frac{1}{2} Dc_i t v_i^2 p(\zeta_i, t), \quad (95)$$

$$\frac{\partial}{\partial t} p(v_i, t) = \zeta_i \frac{\partial}{\partial v_i} p(v_i, t) - \frac{1}{2} Dc_i t v_i^2 p(v_i, t). \quad (96)$$

Here, $a(t) = 1$ and $b(t) = 0$. From the above equations in the steady state, we calculate $p^{st}(\zeta_i, t)$ and $p^{st}(v_i, t)$ as

$$p^{st}(\zeta_i, t) = exp[-\frac{Dc_i t}{2k_i v_i}\frac{\zeta_i v_i^2}{3}],\ p^{st}(v_i, t) = exp[\frac{Dc_i t}{2\zeta_i}\frac{v_i^3}{3}]. \quad (97)$$

The Fourier transforms of the probability densities, $p(\zeta_i, t)$ and $p(v_i, t)$ , are derived as

`

$$p(\zeta_i,t)=q(\zeta_i,t)p^{st}(\zeta_i,t)=\Phi[t-\frac{\zeta_i}{k_iv_i}]p^{st}(\zeta_i,t), \tag{98}$$

$$p(v_i,t)=q(v_i,t)p^{st}(v_i,t)=\Phi[t+v_i/\zeta_i]p^{st}(v_i,t). \tag{99}$$

We calculate $p(\zeta_i,v_i,t)$ from Eqs. (97) and (98) as

$$p(v_i,t)=q(v_i,t)p^{st}(v_i,t)=exp[-\frac{Dc_it^4}{6}\zeta_i^2-\frac{Dc_it^2}{2}v_i^2]. \tag{100}$$

By using the inverse Fourier transforms, $p(x_i,t)$ and $p(v_i,t)$ are, respectively, presented as

$$p(x_i,t)=[2\pi\frac{Dc_it^4}{3}]^{-1/2}exp[-\frac{3}{2Dc_it^4}x_i^2],\ p(v_i,t)=[2\pi Dc_it^2]^{-\frac{1}{2}}exp[-\frac{1}{2Dc_it^2}v_i^2]. \tag{101}$$

The mean-squared deviations for $p(x_i,t)$ and $p(v_i,t)$ are, respectively, calculated as

$$<x_i^2(t)>=\frac{1}{3}Dc_it^4,\ <v_i^2(t)>=Dc_it^2. \tag{102}$$

## 4. Statistical quantities

First of all, we calculate the non-Gaussian parameter, the correlation coefficient, the entropy, and the combined entropy for the displacement and the velocity. First of all, the non-Gaussian parameter for displacement and velocity are, respectively, given by

$$K_x=<x^4>/3<x^2>^2, K_{v_x}=<v_x^4>/3<v_x^2>^2, \tag{103}$$

$$K_y=<y^4>/3<y^2>^2,\ K_{v_y}=<v_y^4>/3<v_y^2>^2. \tag{104}$$

We introduce the correlation coefficient as

$$\rho_{x,v_x}=x_0v_{x0}/\sigma_x\sigma_{v_x},\ \rho_{y,v_y}=y_0v_{y0}/\sigma_y\sigma_{v_y} \tag{105}$$

The entropies $S(x,t)$, $S(v_x,t)$ and $S(y,t)$, $S(v_y,t)$ are, respectively, calculated as

$$S(x,t)=-\int dxf(x,t)lnf(x,t),\ S(v_x,t)=-\int dv_xf(v_x,t)lnf(v_x,t), \tag{106}$$

$$S(y,t)=-\int dyf(y,t)lnf(y,t),\ S(v_y,t)=-\int dv_yf(v_y,t)lnf(v_y,t). \tag{107}$$

and the combined entropy is defined by

$$S(x,v_x,t)=-\int dx\int dv_xf(x,v_x,t)lnf(x,v_x,t)\,, S(y,v_y,t)=-\int dy\int dv_yf(y,v_y,t)lnf(y,v_y,t). \tag{108}$$

Next, in multiplying and integrating both sides of the Fokker–Plank equations by$x^mv^n$, the moment equations for a charged particle from Eq. (7) and Eq. (8) are expressed as

$$\frac{d}{dt}\mu_{m,n}=-m\mu_{m-1,n+1}+\frac{c_i^2}{\omega_i^2}n\mu_{m,n}+Dn(n-1)\mu_{m,n-2}, \tag{109}$$

$$\frac{d}{dt}\overline{\mu}_{m,n}=-m\overline{\mu}_{m-1,n+1}+nk_i\overline{\mu}_{m+1,n-1}-n[c_i(x_0+v_0)t+\frac{c_i^2}{\omega_i^2}(1-t)]\overline{\mu}_{m,n-1}+c_iDtn(n-1)\overline{\mu}_{m,n-2}. \tag{110}$$

Here, $\mu_{m,n}=\int_{-\infty}^{+\infty}dx\int_{-\infty}^{\infty}dv_xx^mv_x^np(x,v_x,t)$, and $\overline{\mu}_{m,n}=\int_{-\infty}^{+\infty}dx_i\int_{-\infty}^{+\infty}dv_ix_i^mv_i^n\,p(x_i,v_i,t)$. The moment equations from Eq. (30) and Eq. (31) for a charged particle with correlated Gaussian forces have

$$\frac{d}{dt}\mu_{m,n}=-m\mu_{m-1,n+1}+\frac{c_i^2}{\omega_i^2}n\mu_{m,n}-mnDb(t)\mu_{m-1,n-1}+mnDa(t)\mu_{m,n-2}, \tag{111}$$

$$\frac{d}{dt}\overline{\mu}_{m,n}=-m\overline{\mu}_{m-1,n+1}+nk_i\overline{\mu}_{m+1,n-1}-n[c_i(x_0+v_0)t+\frac{c_i^2}{\omega_i^2}(1-t)]\overline{\mu}_{m,n-1}-mnDb(t)\mu_{m-1,n-1}+mnDa(t)\mu_{m,n-2}. \tag{112}$$

Table 1 is summarized values of the non-Gaussian parameter, the correlation coefficient, the entropy, and the combined entropy for the motion of a quantum particle and a bath particle with a Gaussian white noise in the short- and long-time domains. Table 2 is summarized values of the non-Gaussian parameter, the correlation coefficient, the entropy, and the combined entropy for the motion of a quantum particle and a bath particle with a correlated Gaussian force within the limits of $t \ll \tau$, $t \gg \tau$ and for $\tau = 0$, where $\tau$ is the correlation time.

**Table 1.** Values of the non-Gaussian parameter, the correlation coefficient, the entropy, and the combined entropy for the motion of a quantum particle and a bath particle with a Gaussian white noise in all times. Here, we assume that quantum particle and bath particle are initially at $x = x_0$, $v = v_0$ and at $x_i = x_{i0}$, $v_i = v_{i0}$, respectively.

| Time | $x, v$<br>$x_i, v_i$ | $K_x, K_v$<br>$K_{x_i}, K_{v_i}$ | $\rho_{x,v}$,<br>$\rho_{x_i,v_i}$ | $\mu_{2,2}$, $\overline{\mu}_{2,2}$ | $S(x,t), S(v,t)$<br>$S(x_i,t), S(v_i,t)$ | $S(x,v,t)$<br>$S(x_i,v_i,t)$ |
|---|---|---|---|---|---|---|
| all times | $x$ | $\frac{x_0^4}{D^2}t^{-6}+\frac{x_0^2}{D}t^{-3}$ | $\frac{x_0v_0}{Dc_i^3/\omega_i^2}t^{-3}$ | $D^2t^4$ | $lnDt^3$ | $\ln(D^2c_i^6/\omega_i^4)t^6$ |
| | $v$ | $\frac{v_0^4}{D^2c_i^{12}/\omega_i^8}t^{-6}+\frac{v_0^2}{Dc_i^6/\omega_i^4}t^{-3}$ | | | $\ln(Dc_i^6/\omega_i^4)t^3$ | |
| | $x_i$ | $\frac{x_{i0}^4}{D^2c_i^2}t^{-8}+\frac{x_{i0}^2}{Dc_i}t^{-4}$ | $\frac{x_{i0}v_{i0}}{Dc_i}t^{-3}$ | $D^2c_i^2t^6$ | $lnDc_it^4$ | $lnD^2c_i^2t^6$ |
| | $v_i$ | $\frac{v_{i0}^4}{D^2c_i^2}t^{-4}+\frac{v_{i0}^2}{Dc_i}t^{-2}$ | | | $lnDc_it^2$ | |

**Table 2.** Values of the non-Gaussian parameter, the correlation coefficient, the entropy, and the combined entropy for the motion of a quantum particle and a bath particle with a correlated Gaussian force within the limits of $t \ll \tau$, $t \gg \tau$ and for $\tau = 0$, where $\tau$ is the correlation time. Here, we assume that quantum particle and bath particle are initially at $x = x_0$, $v = v_0$ and at $x_i = x_{i0}$, $v_i = v_{i0}$, respectively.

| Time | $x, v$<br>$x_i, v_i$ | $K_x, K_v$<br>$K_{x_i}, K_{v_i}$ | $\rho_{x,v}$,<br>$\rho_{x_i,v_i}$ | $\mu_{2,2}$, $\overline{\mu}_{2,2}$ | $S(x,t), S(v,t)$<br>$S(x_i,t), S(v_i,t)$ | $S(x,v,t)$<br>$S(x_i,v_i,t)$ |
|---|---|---|---|---|---|---|
| $t \ll \tau$ | $x$ | $\frac{\tau^2x_0^4}{D^2}t^{-8}+\frac{x_0^2}{D/\tau}t^{-4}$ | $\frac{x_0v_0}{Dc_i^{3/2}/\tau\omega_i}t^{-7/2}$ | $\frac{D^2}{\tau^2}t^6$ | $\ln(D/\tau)t^4$ | $\ln(D^2c_i^3/\tau^2\omega_i^2)t^7$ |
| | $v$ | $\frac{v_0^4}{D^2c_i^6/\tau^2\omega_i^4}t^{-6}+\frac{v_0^2}{Dc_i^3/\tau\omega_i^2}t^{-3}$ | | | $\ln(Dc_i^3/\tau\omega_i^2)t^3$ | |
| | $x_i$ | $\frac{x_{i0}^4}{D^2c_i^2/\tau^2}t^{-10}+\frac{x_{i0}^2}{Dc_i/\tau}t^{-5}$ | $\frac{x_{i0}v_{i0}}{Dc_i(1+\frac{2c_i^2}{\omega_i^2})^{1/2}/\tau}t^{-4}$ | $(D^2c_i^2/\tau)t^7$ | $\ln(\frac{Dc_i}{\tau})t^5$ | $\ln(D^2c_i^2(1+\frac{2c_i^2}{\omega_i^2})/\tau^2)t^8$ |
| | $v_i$ | $\frac{v_{i0}^4}{D^2c_i^2(1+\frac{2c_i^2}{\omega_i^2})^2/\tau^2}t^{-6}+\frac{v_{i0}^2}{Dc_i(1+\frac{2c_i^2}{\omega_i^2})/\tau}t^{-3}$ | | | $\ln[Dc_i(1+\frac{2c_i^2}{\omega_i^2})/\tau]t^3$ | |
| $t \gg \tau$ | $x$ | $\frac{x_0^4}{D^2}t^{-6}+\frac{x_0^2}{D}t^{-3}$ | $\frac{x_0v_0}{Dc_i^{3/2}/\tau^{1/2}\omega_i}t^{-7/2}$ | $D^2t^4$ | $\ln Dt^3$ | $\ln(D^2c_i^3/\tau\omega_i^2)t^6$ |
| | $v$ | $\frac{v_0^4}{D^2c_i^6/\tau^2\omega_i^4}t^{-6}+\frac{v_0^2}{Dc_i^3/\tau\omega_i^2}t^{-3}$ | | | $\ln(\frac{Dc_i^3}{\tau\omega_i^2})\ t^3$ | |
| | $x_i$ | $\frac{x_{i0}^4}{D^2c_i^2}t^{-8}+\frac{x_{i0}^2}{Dc_i}t^{-4}$ | $\frac{x_{i0}v_{i0}}{Dc_i}t^{-3}$ | $D^2c_i^2t^6$ | $lnDc_it^4$ | $\ln D^2c_i^2t^6$ |
| | $v_i$ | $\frac{v_{i0}^4}{D^2c_i^2}t^{-6}+\frac{v_{i0}^2}{Dc_i}t^{-2}$ | | | $\ln Dc_it^2$ | |
| $\tau = 0$ | $x$ | $\frac{x_0^4}{D^2}t^{-6}+\frac{x_0^2}{D}t^{-3}$ | $\frac{x_0v_0}{D}t^{-2}$ | $D^2t^4$ | $\ln Dt^3$ | $\ln D^2t^4$ |
| | $v$ | $\frac{v_0^4}{D^2}t^{-2}+\frac{v_0^2}{D}t^{-1}$ | | | $\ln Dt$ | |

`

| $x_i$ | $\frac{x_{i0}^4}{D^2c_i^2}t^{-8}+\frac{x_{i0}^2}{Dc_i}t^{-4}$ | $\frac{x_{i0}v_{i0}}{Dc_i}t^{-3}$ | $D^2c_i^2t^6$ | $lnDc_it^4$ | $\ln D^2c_i^2t^6$ |
|---|---|---|---|---|---|
| $v_i$ | $\frac{v_{i0}^4}{D^2c_i^2}t^{-6}+\frac{v_{i0}^2}{Dc_i}t^{-2}$ | | | $\ln Dc_it^2$ | |

## 5. Conclusions

We have investigated the dynamics of a quantum particle interacting with a thermal environment described by the Caldeira–Leggett master equation [21,22] with a random noise and a correlated Gaussian noise. By introducing characteristic coordinates and applying Fourier and Laplace transforms, we derived analytic expressions for the joint probability density and obtained the mean square displacement of the particle in different time regimes [23-25].

Our results are as follows. The Caldeira–Leggett model with a white random force leads to a random acceleration process, resulting in the anomalous scaling $\langle x^2(t)\rangle \sim t^3$, while the mean square displacement of a bath particle scales as $< x_i^2 > \sim t^4$. For a correlated Gaussian noise, the mean square displacement and the mean square velocity of a bath particle show the super-diffusive growths of $t^5$ and $t^3$, respectively. In addition, the moment $\overline{\mu}_{2,2}$ scales with $t^6$, for a bath particle with a Gaussian white noise in all times. The moments $\mu_{2,2}$ and $\overline{\mu}_{2,2}$ scale with $t^4$ and $t^6$ in $\tau = 0$, consistent with our calculation result, respectively.

The analysis reveals that the transport properties of the system differ significantly from those of the standard dissipative Caldeira–Leggett dynamics. In particular, the bath particles do not directly feel the thermal noise but is driven by the displacement coordinate of a system. Therefore, the temporal correlations of the noise are effectively averaged through the system dynamics. As a result, the scaling behavior of the bath particle becomes insensitive to the correlation time $\tau$, leading to the same power-law dependences for both $\tau \neq 0$ and $\tau = 0$. The results demonstrate that the absence of dissipation fundamentally modifies the transport mechanism in the Caldeira–Leggett framework. Instead of approaching normal diffusion, the system exhibits persistent super-ballistic quantum spreading driven by correlated environmental noise. The present study provides insight into anomalous transport processes in open quantum systems and may be relevant for understanding nonequilibrium quantum dynamics in correlated environments [26-28].

## References

[1] E.E. Torres-Miyares, G. Rojas-Lorenzo, J. Rubayo-Soneira, and S. Miret-Artes, Surface diffusion within the Caldeira–Leggett formalism, Phys. Chem. Chem. Phys. 24, 15871 (2022).

[2] C. Charalambous, Quantum Brownian Motion in Bose-Einstein Condensates, Ph.D. thesis, Universitat Politècnica de Catalunya (2020).

[3] F. Gottwald, S.D. Ivanov, and O. Kühn, Applicability of the Caldeira–Leggett Model to Vibrational Spectroscopy in Solution, J. Phys. Chem. Lett. 6, 2722 (2015).

[4] U. Weiss, Quantum Dissipative Systems, 4th ed., Series in Modern Condensed Matter Physics Vol. 13 World Scientific, Singapore, 2012).

[5] C.W. Gardiner and P. Zoller, Quantum Noise (Springer, Berlin, 2004).

[6] H.-P. Breuer and F. Petruccione, The Theory of Open Quantum Systems (Oxford University Press, 2002).

[7] D. Kondaurov and E. Polyakov, Quantum Brownian Motion as a Classical Stochastic Process in Phase Space, arXiv:2512.08641v2.

[8] L. Buxton, M.-T. Russo, J. Al-Khalili and A. Rocco, Quantum Brownian Motion in the Caldeira-Leggett Model with a Damped Environment, arXiv:2303.09516v1

[9] A.O. Caldeira and A.J. Leggett, Quantum tunnelling in a dissipative system, Ann. Phys. 149, 374 (1983)

[10] A.O. Caldeira and A.J. Leggett, Path integral approach to quantum Brownian motion, Physica A 121, 587 (1983).

[11] G. Lindblad, On the Generators of Quantum Dynamical Semigroups, Commun. Math. Phys. 48, 119 (1976).

[12] D. Kondaurov and E. Polyakov, Quantum Brownian Motion as a Classical Stochastic Process in Phase Space, arXiv:2512.08641v2.

[13] L. Buxton, M.-T. Russo, J. Al-Khalili and A. Rocco, Quantum Brownian Motion in the Caldeira-Leggett Model with a Damped Environment, arXiv:2303.09516v1.

`

[14] M.A.F. dos Santos and L. Menon, Random diffusivity models for scaled Brownian motion, Chaos, Solitons and Fractals 144, 110634 (2021).
[15] K.S. Olsen, A. Hansen and E.G. Flekkoy, Hyper-Ballistic Superdiffusion of Competing Microswimmers, Entropy 26, 274 (2024).
[16] E.G. Flekkøy, A. Hansen and B. Baldelli, Hyperballistic superdiffusion and explosive solutions to the non-linear diffusion equation, arXiv:2302.13666v6.
[17] S. Aerdker, L. Merten, F. Effenberger, H. Fichtner and J.B. Tjus, Superdiffusion of energetic particles at shocks: A Levy Flight model for acceleration, arXiv:2408.03638v1.
[18] H. Ikeda, Correlated Noise and Critical Dimensions, arXiv:2302.13666v6.
[19] W. Wang, Q. Wei, A.V. Chechkin and R. Metzler, Different behaviors of diffusing diffusivity dynamics based on three different definitions of fractional Brownian motion, arXiv:2504.19190v1.
[20] J. Heinrichs, Probability distributions for second-order processes deriven by Gaussian noise. Phys. Rev. E 47, 3007 (1993).
[21] A.J. Leggett, Quantum tunneling in the presence of an arbitrary linear dissipation mechanism, Phys. Rev. B 30, 1208 (1984).
[22] C.W. Gardiner and P. Zoller, Quantum Noise (Springer, Berlin, 2004)
[23] J.W. Jung, S.K. Seo and K. Kim, Joint probability densities of an active particle coupled to two heat reservoirs Physica A 668, 130483 (2025).
[24] J.W. Jung, S.K. Seo and K. Kim, Dynamical behavior of passive particles with harmonic, viscous, and correlated Gaussian forces Phys. Lett. A 546 130512 (2025).
[25] Y.J. Kang, J.W. Jung, S.K. Seo and K. Kim, On the Stochastic Motion Induced by Magnetic Fields in Random Environments Entropy 27, 330 (2025).
[26] K. Luoma, W.T. Strunz and J. Piilo, Diffusive Limit of Non-Markovian Quantum Jumps Phys. Rev. Lett. 125, 150403 (2020).
[27] X. Zhang, Z. Wu, G.A.L. White et al., Learning and forecasting open quantum dynamics with correlated noise, Commun. Phys. 8, 29 (2025).
[28] D. Gamba, B. Cui and A. Zaccone, Open quantum systems with particle and bath driven by time-dependent fields, arXiv:2505.05348v2.